\documentclass[12pt]{article}
\usepackage{graphicx}
\usepackage{amsmath}
\usepackage[labelformat=simple]{subfig} 

\begin{document}

\title{FastGCN+ARSRGemb: a novel framework for object recognition}


\author{Mario Manzo$^1$ \and Simone Pellino$^2$}
\date{%
    $^1$ Information Technology Services\\ 
	University of Naples “L’Orientale”\\
              80121, Naples, Italy\\
              mmanzo@unior.it \\%
    $^2$Enrico Mattei Istitute\\
           581031, Aversa (CE), Italy\\
           simonepellino@gmail.com\\%
}

\maketitle

\begin{abstract}
In recent years research has been producing an important effort to encode the digital image content. Most of the adopted paradigms only focus on local features and lack in information about location and relationships between them. To fill this gap, we propose a framework built on three cornerstones. First, ARSRG (Attributed Relational SIFT (Scale-Invariant Feature Transform) regions graph), for image representation, is adopted. Second, a graph embedding model, with purpose to work in a simplified vector space, is applied. Finally, Fast Graph Convolutional Networks perform classification phase on a graph based dataset representation. The framework is evaluated on state of art object recognition datasets through a wide experimental phase and is compared with well-known competitors.
\end{abstract}


\section{Introduction}
\label{intro}
In the last few years the image classification has benefited from a wide range of methods in which the features are extracted and stored in a descriptive format for an easy and correct access by multimedia tools. In many cases, the main problem concerns the discrimination and importance of details to include in the image description and the related computational cost which inevitably leads to a degradation of performance. As result, the main challenge concerns the discriminating power of the features combined with the optimization of the performance. Starting from these drawbacks, our goals refer to features (local and global) to include in the image representation, in order to improve time processing, and the creation of global connections between images to derive similarities based on their content.
In this regard, we propose a framework, called \textbf{FastGCN+ARSRGemb}, composed of three modules. First, image representation. We encode both spatial and structural information adopting a graph based representation termed ARSRG (Attributed Relational SIFT (Scale-Invariant Feature Transform) Regions Graph) \cite{manzo2013attributed}. The second concerns graph embedding. The images, in ARSRG format, are transformed into vector space through a graph embedding procedure. Third, a combination of a global graph-based representation and FastGCN (Fast Graph Convolutional Networks). Each image is represented, from a global point of view, as a point connected/disconnected to images, based on a similarity criterion, present in the same set. The classification phase is managed with FastGCN application on resulting graph. The framework is adapted within an object recognition context in which the main critical issues involve best image representation and matching, which could affect the quality of performance. The paper is organized as follows: in section \ref{relatedwork} object recognition approaches are analyzed. In section \ref{framework} the proposed framework is described. Section \ref{res} provides a wide experimental phase. Finally section \ref{conc} reports conclusions and future works.

\section{Related work}
\label{relatedwork}
 
Finding and identifying objects in a video or image sequence is a very difficult task. Starting from a set of known models, the goal is to assign correct labels to the regions containing objects of interest. The main critical issues are related to the representations and classification. The aim is to emulate the human system, which performs efficiently and dynamically the task.

In \cite{wang2006tensor} authors provide an $M^{th}$ order tensor discriminant analysis
approach for object categorization and recognition. The method represents a color image as a $M^{th}$ order tensor mapped into a low dimensional features space. Nearest neighbor classifiers and AdaBoost (from now DTROD-AdaBoost) are employed to perform the final classification.

In \cite{obdrzalek2002object} the goal concerns several affine-invariant constructions of Local Affine Frames (LAFs) for local image patches extraction. The matching procedure by selecting multiple frames to each image region is performed. A number of established local correspondences for matching is considered instead of global model consistency.

In \cite{wang2006tensor,obdrzalek2002object} location and spatial information between local features are not included. In order to reduce this gap, graph structures can be adopted. Graphs appear in application domains, image processing \cite{manzo2013attributed,acosta2012frequent}, pattern recognition \cite{Rozza:kernel,gago2010full,jia2011efficient}, and others, where highlight relations among data is essential. The literature provides several solutions despite Region Adjacency Graph (RAG) \cite{Tremeau:regions}, in which regions and spatial relations between regions as nodes and edges are encoded respectively, is the most used.

In \cite{HoriTakiguchiAriki} graph structural expression model for generic object recognition is proposed. This model, at same time, creates a graph structure for local features connection with purpose to reduces the computation complexity and improves the detection performance.

In \cite{gago2010full} a graph mining algorithm, called gdFil, is introduced. Two novel properties allowing to remove all duplicate candidates in Frequent Connected Subgraph (FCS) before support calculation are explored. Support calculation task is addressed through a strategy based on embedding structures. 

In \cite{jia2011efficient} authors propose a graph mining framework called APproximate Graph Mining (APGM). The framework identifies approximate matched FCSs and mines useful patterns from noisy graph database.

In \cite{acosta2012frequent} Vertex and Edge Approximate graph Miner (VEAM), a graph mining algorithm for FCSs, is described. VEAM works both on vertex and edge label sets during the mining process. Experimental results in the context of graph-based image classification are produced.

In \cite{XiaHancock} an object is represented adopting local features selected by a model based on visual saliency. Specifically, the objects are built by a Class Specific Hyper-Graphs (CSHG) using Delaunay graphs which include local features as nodes. 

In \cite{morales2014new} an image by an irregular pyramid is represented, where each level is a RAG and the base level is the entire image. In order to enrich the structure, image regions are represented by different basic low-level descriptors and by employing Frequent Approximate Subgraph (FAS).

In \cite{morales2013simple} authors adopt visual features (color, texture and shape) and spatial relations (VFSR) to detect similar objects. The proposed combination can be adopted to encode many possible spatial configurations among image regions, also with different orientation and topological relationships.

In \cite{maree2005decision} hierarchical features, to capture local and structural information about object, are adopted (from now RSW+Boosting). The method combines decision trees for classification task.

In \cite{manzo2019bag} a paradigm called Bag of ARSRG (Attributed Relational SIFT (Scale-Invariant Feature Transform) Regions Graph) Words (BoAW) is proposed. A digital image is described as a vector in terms of a frequency histogram of graphs. Adopting different steps, the images are mapped into a vector space passing through a graph transformation. Also in \cite{manzo2014novel} ARSRG is adopted to build a feature vector computed by means of a graph embedding paradigm. 

Literature provides additional models which adopt local and spatial information different to graph structures. One of these is in \cite{morioka2008learning}, where a temporal model based on local features (from now Sequential Patterns) is adopted. The temporal information with spatial relations and local features is combined, in order to manage recognition task as a sequential prediction task. 

\section{Framework overview}
\label{framework}

In this section, we describe the proposed object recognition framework, named \textbf{FastGCN} + \textbf{ARSRGemb}, composed by different modules. In the first module, each image is encoded through an Attributed Relational SIFT-based Graph (ARSRG \cite{manzo2013attributed}). This structure is able to capture both local and spatial information of image. Second module maps each ARSRG into a vector space through a graph embedding paradigm. Each component of the vector includes the distance, computed by an efficient graph matching algorithm \cite{manzo2013attributed}, between ARSRGs. The third module creates a global graph. Nodes represent the individual ARSRG and the edges the connections between them. Finally, Fast Graph Convolutional Networks (FastGCN) is trained on the on the global graph with purpose to address the classification task. 

\subsection{Graph Based Image Representation}
\label{imagetograph}

Attributed Relational SIFT-based Regions Graph (ARSRG), proposed in \cite{manzo2013attributed}, is employed to described image content. The structure is
composed by three different levels of nodes: the Root node,
the RAG Nodes, and the Leaf nodes. The
Root node represents the whole image and is linked with all
the RAG Nodes \cite{tremeau2000regions} of the second
level. RAG Nodes represent image regions, extracted
by means of a segmentation technique, and encode adjacency
relationships between them. At this level, adjacent regions in the
image are represented by connected nodes. Finally, the Leaf
nodes represent the set of SIFT \cite{Lowe} descriptors
extracted from the image. Precisely, a descriptor is associated to a region based on its
spatial coordinates and the descriptors belonging to the same region
are connected by edges. SIFT guarantee invariance to the view-point, illumination and scale. Compared to the approach proposed in~\cite{manzo2013attributed}, in which the ARSRG structures are adopted to solve an image retrieval problem, we employ ARSRG structures to map image features through the embedding procedure described in the section \ref{graphembedding}.

\subsection{Graph embedding}
\label{graphembedding}

Literature provides many approaches for dimensionality reduction such as Principal Component Analysis (PCA) \cite{Jollife}, Linear Discriminant Analysis (LDA), \cite{Bishop06}, and Kernel variants of this techniques \cite{Bishop06}. 
The main goal is to produce lower dimensional representation from the original higher dimensional features space preserving some properties of the data. Differently, our goal is to work with structured data. To this end, we adopt a graph embedding model \cite{manzo2014novel} with the purpose to provide a fixed-dimensional vector representation of ARSRG structures. Given a set of sample labeled graphs
$S=\{G_{1},\ldots,G_{n}\}$ and a graph similarity
measure $s(G_{i},G_{j})$, where $S$ and $s(G_{i},G_{j})$ can be any kind of graph set and graph similarity measure respectively. Moreover, given a set
$P=\{P_{1},\ldots,P_{m}\}$ of $m=n$ prototypes
extracted from $S$, the aim is to compute the similarities of a given input graph $G_j$ with each prototype $P_{k} \in P$. The procedure produces $m$ similarities,
$s_{1}=s(G_j,P_{1}),\ldots,s_{m}=s(G_j,P_{m})$,
in form of $m$-dimensional vector
$(s_{1},\ldots,s_{m})$. In this way, any graph can be transformed into a vector of real numbers. Summing up, if we consider a graph domain $G$, the training set of graphs $S=\{G_{1},\ldots,G_{n}\} \subseteq G$, and
a set of prototype graphs $P=\{P_{1},\ldots,P_{m}\} \subseteq S$, the vector of mapping between $S$ and $P$ is defined as follows:
\begin{equation}
\label{gremb}
\Phi_{m}^{P}(G_m)=(s(G_m,P_{1}),\ldots,s(G_m,P_{m}))
\end{equation}%
where $s(G_m,P_{i})$ is a graph similarity measure
between graph $G_m$ and the $i$th prototype. We apply this paradigm to obtain a vector where components encode the distance, obtained through an iterative and efficient graph matching algorithm \cite{manzo2013attributed}, between the considered ARSRGs and related prototypes. Specifically, regions similarities among the ARSRGs are measured through the exploration of topological relations. The algorithm works on two levels. The first explores global features, regions extracted through JSEG segmentation \cite{deng2001unsupervised}, while second explores local invariant region features. In this way, both local and structural information during the matching process are analyzed.

\subsection{Data graph representation}
\label{graphrep}

Graph embedding result is composed by multidimensional vectors. At this step, the goal is to provide a graph representation, $G_{ARSRGs}$ from now, in order to encode the ARSRGs set as a multidimensional space points. $G_{ARSRGs}$ can be defined as:

\begin{equation}
G_{ARSRGs}=(V,E,X,Y)    
\end{equation}

where 
\begin{equation}
    V=(ARSRG_{1},\ldots,ARSRG_{n})
\end{equation} 

is the set of nodes which represent ARSRGs, while 

\begin{equation}
E \subseteq \{\{i,j\} | (i,j) \in V^2 \land i \ne j\}, 
\end{equation}

is the set of edges among ARSRGs. Additionally, matrix 
\begin{equation}
\label{featmat}
    X=(\Phi_{m}^{P}(ARSRG_1),\ldots,\Phi_{m}^{P}(ARSRG_n))
\end{equation} 

refers to equation \ref{gremb} and represents node features, in form of $\Phi$ vectors. Finally, 

\begin{equation}
\label{nodelabels}
 Y=(l_1,\ldots,l_n)   
\end{equation}

includes node labels, $ARSRGs$ membership class, useful for classification stage. Now, it is important to define the adjacency matrix $A$ of $G_{ARSRGs}$, representing the connections among ARSRGs, as follows:

\begin{equation}
\label{adj}
  A_{i,j}=\begin{cases}
    1, & \text{if $s(ARSRG_{i},ARSRG_{j}) < \tau$}.\\
    0, & \text{otherwise}.
  \end{cases}
\end{equation}

threshold $\tau$ gives the sensitivity to accept a certain distance between ARSRGs. $A_{i,j}=1$ indicates that $ARSRG_{i}$ and  $ARSRG_{j}$ are connected (similar). The diagonal elements $A_{i,i}$ are set to 0 to avoid selfsimilarity. 

\subsection{Fast Graph Convolutional Networks}
\label{GCN}

Graph Convolutional Networks (GCN) \cite{kipf2016semi} is the architecture that applies a convolution on graphs . The main concept concerns the convolution filter, originally applied to image pixels or a linear matrix of signals. GCN focus on the topology of the graph as filter to perform neighborhood exploration. The architecture is enclosed in the following equation:

\begin{equation}
    H^{(l+1)}=\sigma(\hat{A} H^{(l)} W^{(l)}),
\end{equation}

$\hat{A}$ represents a normalization of the graph adjacency matrix, defined in equation \ref{adj}, $H^{(l)}$ represents the embedding, based on rows, of the graph vertices in the $l$th layer, with $H^{(0)}= X$ and $X$ defined in equation \ref{featmat}, $W^{(l)}$
is a parameter matrix, and $\sigma$ is nonlinearity. In particular, the adjacency matrix describes the relationship in pairs for both training data and tests. Learning and embedding are performed in parallel for both data. Frequently, test data may not be readily available, since the graph can constantly increase with additional nodes. These cases need a learning process only from a training set and a generalization of the growing of graph. A thorny issue for GCN concerns the recursive exploration of neighborhoods across layers involving expensive calculations in batch training. Especially in high density and powerlaw cases, exploring the neighborhood for a single node invades rapidly  a large part of the graph. Therefore, scalability is the main problem for GCN processing on big and dense graphs. In \cite{chen2018fastgcn} a fast version of GCN, called Fast Graph Convolutional Networks (FastGCN), is proposed. It processes graph convolutions differently and manage them as integral transformations of the embedding functions in probability measures to reduce both aforementioned gaps. This approach provides a principle mechanism for inductive learning, reformulating the loss as stochastic version of the gradient. In particular, the vertices of the graphs represent iid samples of a probability distribution. While, the loss of each convolution layer is written as integral related to the vertex embedding function. In our case FastGCN through the graph defined in section \ref{graphrep} is adopted. Particularly, it uses adjacency matrix $A$, equation \ref{adj}, node features matrix $X$, equation \ref{featmat}, and node label vector $Y$, equation \ref{nodelabels}.

\section{Experimental results}
\label{res}

This section describes the experiments performed on public datasets. In order to produce compliant performance, the settings described in well-known object recognition methods, in which the features selection to best represent objects-classes is the main critical issue, are adopted.

\subsection{Datasets}
\label{datasets}

\textbf{FastGCN} + \textbf{ARSRGemb} on a dataset containing object images is tested. Datasets adopted are: \begin{enumerate}

           \item Amsterdam Library of Object Images (ALOI) \cite{geusebroek2005amsterdam}. It is a color image selection of 1000 small objects. Objects was recorded varying viewing angle, illumination angle and illumination color in order to capture the sensory variation. In addition was captured wide-baseline stereo image.
            \item Columbia Object Image Library (COIL-100) \cite{nene1996object}.  It is a image collection of 100 objects. The images were taken at pose internals of 5 degrees. 
                    \item The ETH-80 \cite{leibe2003analyzing}. It is a image collection of 80 objects from 8 categories. Each object is described by 41 different views, thus obtaining a total of $3280$ images.

\end{enumerate}

\begin{figure}[!ht]
\centering
\subfloat[]{\includegraphics[width=0.2\textwidth]{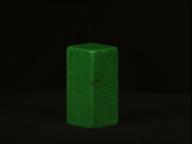}
    }\qquad
\subfloat[]{\includegraphics[width=0.15\textwidth]{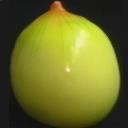}
    }\\
\subfloat[]{\includegraphics[width=0.2\textwidth]{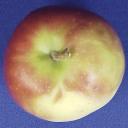}
    }%
    
\caption{Dataset images: (\textbf{a}) ALOI, (\textbf{b}) COIL-100, (\textbf{c}) ETH-80.} \label{ds}
\end{figure}

Figure \ref{ds} shows some examples of datasets.

\subsection{Experimental Setup}

One-versus-All (OvA) paradigm for experimental procedure is adopted. Images, before being converted to ARSRG, are scaled ($150 \times 150$ pixels size) in order to reduce execution time. Image segmentation parameters are: threshold for color quantization is in the range $0-600$, threshold for the region merging is fixed to $0.4$. The SIFT dimensional vector is fixed to $128$. Settings for graph matching are: threshold for false positives is fixed to $0.6$ and the minimum number of SIFTs, for region matching, is fixed to $3$. For FastGCN, different configurations, reported in table \ref{FastGCNset}, are adopted to optimize performance based on dataset. First of all, ETH-80 needed a greater number of epochs because the images do not have a uniform background and, therefore, FastGCN takes more time during the learning phase. This behavior is related to the ARSRG structure which cannot always isolate the background from the object, including unnecessary information into image description. Consequently, this also affects the recognition phase. The smartest configuration, allowing a lower execution time, is foreseen for ALOI with a number of hidden units equal to 128. Furthermore, a lower $\tau$ value is adopted for COIL-100 which enables to select a smaller amount of edges making the graph more sparse compared to ETH-80 and COIL-100. This setup favors the processing, as explained in section \ref{GCN}, since one of the limits of GCN regards a slowdown with high density graphs.

\begin{table}[ht!]
\caption{FastGCN settings.}
\label{FastGCNset}
\centering \tabcolsep=0.11cm
\tiny{
\begin{tabular}{cccccccc}
\hline\noalign{\smallskip}
\textbf{Dataset} & \textbf{epochs} & \textbf{hidden size} & \textbf{learning rate} & \textbf{l2 regularization} & \textbf{batch size} & \textbf{sample size} & \textbf{$\tau$}\\ 
\hline\noalign{\smallskip}
ETH-80 & 30000 & 256 & 0.1 & 0 & 1024 & $\frac{|V|}{2}$ & 0.2\\ 
\hline\noalign{\smallskip}
COIL-100 & 10000 & 512 & 0.1 & 0 & 1024 & $\frac{|V|}{2}$ & 0.1\\ 
\hline\noalign{\smallskip}
ALOI & 5000 & 128 & 0.1 & 0 & 256 & $\frac{|V|}{2}$ & 0.2\\ 
\hline\noalign{\smallskip}
\end{tabular}
}
\end{table}

The framework is composed of Matlab code. Moreover, we adopted the related code, in C language, of the JSEG algorithm \cite{deng2001unsupervised}, and the related code, a combination of Matlab and C language, of the SIFT~\cite{Lowe} algorithm.

\subsection{Discussion}
\label{discussion}

Table \ref{ALOI} reports accuracy on the ALOI dataset and settings described in \cite{uray2007incremental}. Specifically, only the results considering a batch of 400 images are considered, since intermediate results do not provide particular improvements. We show the results achieved by BoAW \cite{manzo2019bag}, BoVW \cite{lazebnik2006beyond} and those obtained in \cite{uray2007incremental} using some variants of linear discriminant analysis (ILDAaPCA, batchLDA, ILDAonK and ILDAonL) and ARSRGemb \cite{manzo2014novel}.  Results are listed in form of accuracy. As can be seen, \textbf{FastGCN} + \textbf{ARSRGemb} provides best performance for the object recognition task and there is no degradation when the number of images increases. Indeed, the combination of local and spatial information provides clear benefits in image representation and matching. Furthermore, the graph representation $G_{ARSRGs}$, introduced in section \ref{graphrep}, encodes images, belonging to the same set, in term of distances, providing additional information to improve the image description. Last but not least aspect, connected to two factors:  ARSRGs building and FastGCN learning, concerns the increase in processing time when the images considered grow, in the range 200-3600. Both steps of the pipeline take longer to process. Despite this, in all reported cases the performance remains unchanged.

\begin{table}[ht!]
\caption{Results on the ALOI 
	 dataset.}
\label{ALOI}
\centering \tabcolsep=0.11cm
\tiny
\resizebox{\columnwidth}{!}{%
\begin{tabular}{ccccccccccc}
\hline\noalign{\smallskip}
\textbf{Method} & \textbf{200} & \textbf{400} & \textbf{800} & \textbf{1200} & \textbf{1600} & \textbf{2000} & \textbf{2400} & \textbf{2800} & \textbf{3200} & \textbf{3600}\\
\hline\noalign{\smallskip}
\textbf{FastGCN} + \textbf{ARSRGemb} & \textbf{99.60}\% & \textbf{99.60}\% & \textbf{99.80}\% & \textbf{99.65}\% & \textbf{99.90}\% & \textbf{99.10}\% & \textbf{99.00}\% & \textbf{99.80}\% & \textbf{99.20}\% & \textbf{99.30}\%\\
\hline\noalign{\smallskip}
BoAW & 98.29\% & 92.83\% & 98.80\% & 96.80\% & 96.76\% & 98.15\% & 89.52\% & 82.65\% & 79.96\% & 79.88\%\\
\hline\noalign{\smallskip}
ARSRGemb & 86.00\% & 90.00\%  & 93.00\%  & 96.00\% & 95.62\% & 96.00\% & 88.00\% & 81.89\% & 79.17\% & 79.78\%\\
\hline\noalign{\smallskip}
BoVW & 49.60\% & 55.00\% & 50.42\% & 50.13\% & 49.81\% & 48.88\% & 49.52\% & 49.65\% & 48.96\% & 49.10\%\\
\hline\noalign{\smallskip}
batchLDA  & 51.00\% & 52.00\% & 62.00\% & 62.00\% & 70.00\% & 71.00\% & 74.00\% & 75.00\% & 75.00\% & 77.00\%\\
\hline\noalign{\smallskip}
ILDAaPCA  & 51.00\% & 42.00\% & 53.00\% & 48.00\% & 45.00\% & 50.00\% & 51.00\% & 49.00\% & 49.00\% & 50.00\%\\
\hline\noalign{\smallskip}
ILDAonK  & 42.00\% & 45.00\% & 53.00\% & 48.00\% & 45.00\% & 51.00\% & 51.00\% & 49.00\% & 49.00\% & 50.00\%\\
\hline\noalign{\smallskip}
ILDAonL  & 51.00\% & 52.00\% & 61.00\% & 61.00\% & 65.00\% & 69.00\% & 71.00\% & 70.00\% & 71.00\% & 72.00\%\\
\hline\noalign{\smallskip}
\end{tabular}
}
\end{table}

Table \ref{COIL} shows results on the COIL-100 dataset. In order to obtain a valid comparison with the methods in \cite{morales2014new,morales2013simple} we adopted the same settings: 25 objects are randomly selected and 11\% are used as the training set and 89\% are used as the testing set. Therefore, we compared with BoAW \cite{manzo2019bag}, BoVW \cite{lazebnik2006beyond},  VFSR \cite{morales2014new,morales2013simple}, gdFil \cite{gago2010full}, APGM \cite{jia2011efficient}, VEAM \cite{acosta2012frequent}, DTROD-AdaBoost \cite{wang2006tensor}, RSW+Boosting \cite{maree2005decision} , Sequential Patterns \cite{morioka2008learning}, LAF \cite{obdrzalek2002object}  and ARSRGemb \cite{manzo2014novel}. Results are listed in form of accuracy. Also in this case our approach confirms its qualities obtaining the best performance.

\begin{table}[ht!]
\caption{Results on the COIL-100 dataset.}
\label{COIL}
\centering \tabcolsep=0.11cm
\tiny
\begin{tabular}{cc}
\hline\noalign{\smallskip}
\textbf{Method} & \textbf{Accuracy}\\
\hline\noalign{\smallskip}
\textbf{FastGCN} + \textbf{ARSRGemb} & \textbf{99.88}\%\\
\hline\noalign{\smallskip}
BoAW & 99.77\%\\
\hline\noalign{\smallskip}
ARSRGemb & 99.55\%\\
\hline\noalign{\smallskip}
BoVW & 51.71\%\\
\hline\noalign{\smallskip}
gdFil & 32.61\%\\
\hline\noalign{\smallskip}
VFSR& 91.60\%\\
\hline\noalign{\smallskip}
APGM & 99.11\%\\
\hline\noalign{\smallskip}
VEAM & 99.44\%\\
\hline\noalign{\smallskip}
DTROD-AdaBoost  & 84.50\%\\
\hline\noalign{\smallskip}
RSW+Boosting & 89.20\%\\
\hline\noalign{\smallskip}
Sequential Patterns  & 89.80\%\\
\hline\noalign{\smallskip}
LAF & 99.40\%\\
\hline\noalign{\smallskip}
\end{tabular}
\end{table}

Table \ref{ETH-80} shows results on the ETH-80 dataset. The same setup reported in \cite{morales2014new} to perform a direct comparison is adopted. The setting consists of six categories (apples, cars, cows, cups, horses, and tomatoes). The training set is composed of $4$ objects for each class and $10$ different views for each object with an amount of $240$ images. The testing set is composed of $60$ images for each category ($15$ views per object). We present tests performed by BoAW \cite{manzo2019bag}, ARSRGemb \cite{manzo2014novel}, BoVW \cite{lazebnik2006beyond}, gdFil \cite{gago2010full}, APGM \cite{jia2011efficient}, VEAM \cite{acosta2012frequent}. Also in this case the results are listed highlighting the accuracy of the best approach. As can be seen \textbf{FastGCN} + \textbf{ARSRGemb} provides better results than competitors also when view points changes occur.

\begin{table}[ht!]
\caption{Results on the ETH-80 dataset.}
\label{ETH-80}
\centering \tabcolsep=0.11cm
\tiny
\begin{tabular}{cc}
\hline\noalign{\smallskip}
\textbf{Method} & \textbf{Accuracy}\\
\hline\noalign{\smallskip}
\textbf{FastGCN} + \textbf{ARSRGemb} & \textbf{97.01}\%\\
\hline\noalign{\smallskip}
BoAW & 89.29\%\\
\hline\noalign{\smallskip}
ARSRGemb & 89.26\%\\
\hline\noalign{\smallskip}
BoVW & 58.83\%\\
\hline\noalign{\smallskip}
{gdFil}& 47.59\%\\
\hline\noalign{\smallskip}
{APGM} & 84.39\%\\
\hline\noalign{\smallskip}
{VEAM} & 82.68\%\\
\hline\noalign{\smallskip}
\end{tabular}
\end{table}

\section{Conclusions and Future Works}
\label{conc}

In this work, a novel framework for object recognition is described. We propose an image classification approach which works to more complex structures including spatial information and local feature points. The goal is to optimize performance through a selected set of image features and a graph based representation for FastGCN processing. The ARSRG representation provides a massive filtering on features and the graph, successively created on ARSRGs set, constitutes the dataset for the classification phase. We produced a wide experimental phase, in which framework assumed a robust behavior when objects present uniform backgrounds (ALOI and COIL-100 datasets) while it behaves differently with uneven background (ETH-80 dataset), with a consequent greater number of epochs for FastGCN learning. Future works concern the improvement of image description with purpose to address different type of image classification/retrieval task.

\section*{Acknowledgements}
Our thinking is for Alfredo Petrosino. He followed us during first steps towards the Computer Science, through a whirlwind of goals, ideas and, specially, love and passion for the work. We will be forever grateful great master.

\bibliographystyle{abbrv}
\bibliography{simple}

\end{document}